\documentclass[aps,prl,twocolumn,showpacs,groupedaddress]{revtex4}  
\usepackage{graphicx}  
\usepackage{dcolumn}   
\usepackage{bm}        
\usepackage{amssymb}   
\newcommand{\dst}{${D}^{*}\:$}

\newcommand{\ds}{${D}_{s1}^{\pm}(2536)\:$}
\newcommand{\bdec}{${B}^0_{s} \rightarrow {D}_{s1}^-(2536) \mu^+ 
\nu_{\mu} {X} \:$}
\newcommand{\brbdec}{$Br({B}^0_{s} \rightarrow
{D}_{s1}^-(2536) \mu^+ \nu_{\mu} {X}) \:$}

\newcommand{\ddec}{${D}_{s1}^{-}(2536) \rightarrow {D}^{* -} K^0_S \:$}
\def\d0{D\O}

\newcommand{\adzero}     {$\bar{D^0}$}

\begin{document}


\hspace{5.2in} \mbox{Fermilab-Pub-07/659-E}

\title{Measurement of the 
\boldmath $B^0_s$ \unboldmath 
semileptonic branching ratio to an orbitally excited 
\boldmath $D_s^{**}$ \unboldmath state, 
\boldmath \brbdec \unboldmath}
%
\author{V.M.~Abazov$^{36}$}
\author{B.~Abbott$^{76}$}
\author{M.~Abolins$^{66}$}
\author{B.S.~Acharya$^{29}$}
\author{M.~Adams$^{52}$}
\author{T.~Adams$^{50}$}
\author{E.~Aguilo$^{6}$}
\author{S.H.~Ahn$^{31}$}
\author{M.~Ahsan$^{60}$}
\author{G.D.~Alexeev$^{36}$}
% ...

\affiliation{\vspace{0.1 in}(The D\O\ Collaboration)\vspace{0.1 in}}
% ...
\affiliation{$^{6}$University of Alberta, Edmonton, Alberta, Canada,
                Simon Fraser University, Burnaby, British Columbia, Canada,
                York University, Toronto, Ontario, Canada, and
                McGill University, Montreal, Quebec, Canada}
% ...
\affiliation{$^{29}$Tata Institute of Fundamental Research, Mumbai, India}
% ...
\affiliation{$^{31}$Korea Detector Laboratory, Korea University, Seoul, Korea}
% ...
\affiliation{$^{36}$Joint Institute for Nuclear Research, Dubna, Russia}
% ...
\affiliation{$^{50}$Florida State University, Tallahassee, Florida 32306, USA}
% ...
\affiliation{$^{52}$University of Illinois at Chicago,
                Chicago, Illinois 60607, USA}
% ...
\affiliation{$^{60}$Kansas State University, Manhattan, Kansas 66506, USA}
% ...
\affiliation{$^{66}$Michigan State University,
                East Lansing, Michigan 48824, USA}
% ...
\affiliation{$^{76}$University of Oklahoma, Norman, Oklahoma 73019, USA}
% ...
\date{February 3, 2009}

\begin{abstract}
In a data sample of approximately 1.3 fb$^{-1}$ collected
with the D0 detector between
2002 and 2006, the orbitally excited charm state \ds has been
observed with a measured mass of
$2535.7 \pm 0.6 \thinspace {\mathrm{(stat)}}
                       \pm 0.5 \thinspace {\mathrm{(syst)}} \thinspace {\mathrm{MeV}}/c^2$
via the decay mode \bdec. 
A first measurement is made of the branching ratio product
$Br(\bar{b} \rightarrow
{D}_{s1}^{-}(2536) \mu^+ \nu_{\mu} {X})
\cdot Br(D_{s1}^- \rightarrow D^{*-} K^0_S).$
Assuming that $D_{s1}^-(2536)$ production in semileptonic
decay is entirely
from $B^0_s$,
an extraction of the semileptonic
branching ratio \brbdec is made. 
\end{abstract}

\maketitle 

Semileptonic $B_s^0$ decays into orbitally excited $P$-wave strange-charm
mesons ($D^{**}_s$) are expected to
make up a significant fraction of $B^0_s$ 
semileptonic decays
and are therefore important when comparing inclusive and exclusive decay
rates, extracting CKM matrix elements, and using semileptonic decays
in $B^0_s$ mixing analyses. 
For $B$ meson
semileptonic decays to heavier excited charm states, more of
the available phase space is near zero recoil, increasing
the importance of corrections in heavy-quark effective theory 
(HQET)~\cite{pred1}, effectively tested here.

$D^{**}_s$ mesons (also denoted $D_{sJ}$) are composed of
a heavy charm quark and a lighter strange quark in an 
$L=1$ state of orbital momentum.
In the heavy-quark limit,
the spin $s_Q$ of the heavy quark and the total angular momentum,
$j_q = s_q + L$ of the light degrees of freedom (quark and gluons),
are separately conserved and the latter has possible values of
$j_q = \frac{1}{2}$ or $\frac{3}{2}$.
The surprisingly light masses of the
$j_q = \frac{1}{2}$ states: $D^*_{s0}(2317)$ and
$D_{s1}(2460)$~\cite{dsstst}, plus the observation of new
$D_{sJ}$ states~\cite{SELEX},
deepens the need
for a better understanding of these $D_s^{**}$ systems
since they may be quark molecular states, a new and very 
different arrangement of quarks.

In our decay of interest,
the $j_q=\frac{3}{2}$ angular momentum 
can combine with the 
heavy quark spin to 
form the  $J^P=1^+$ ($D_{s1}$) state 
which must decay 
through a $D$-wave 
to conserve $j_q=\frac{3}{2}$.
The \ds ~is expected to decay dominantly into a \dst
and $K$ meson 
to conserve angular momentum.  
In this Letter we present the first measurement of 
semileptonic $B^0_s$ decay into the narrow 
$D_{s1}^{\pm}(2536)$ state.  This state is just above the \dst  $K_S^0$ ~mass threshold and
has been observed previously~\cite{past_dsstst}. Events 
compatible with the decay chain
$\bar{b} \rightarrow D_{s1}^-(2536) \mu^+ \nu_{\mu} X, \thinspace
D_{s1}^-(2536) \rightarrow D^{*-} K^0_S; \thinspace
D^{*-} \rightarrow \bar{D}^0 \pi^-, \thinspace
K^0_S \rightarrow \pi^+ \pi^-, 
\thinspace \bar{D}^0 \rightarrow K^+ \pi^- $
%
are reconstructed.
Charge conjugate modes and reactions are always implied in this
Letter. 

Assuming that $D_{s1}^{-}(2536)$ production in a semileptonic decay is entirely from $B_s^0$, the branching ratio \brbdec can be determined by 
normalizing to the known value of the branching fraction
$Br(\bar{b} \rightarrow D^{*-} \mu^+ \nu_{\mu} X) =
(2.75 \pm 0.19)\%$~\cite{PDG} to avoid uncertainties in the
$b$-quark production rate.
This semileptonic branching ratio includes any 
decay channel or sequence of channels resulting in a
$D^*$ and a lepton (muon in our case), and all $b$ hadrons, and
therefore includes the relative production of each $b$ hadron
species starting from a $\bar{b}$ quark.
Since the final state of interest, \ddec, is reconstructed from a $D^*$ and a $K^0_S$, the selection
is broken up into two sections: one to reconstruct the $D^*$ with an 
associated muon, coming dominantly from $B$ meson decays
resulting in a number of candidates, $N_{D^* \mu}$, and
then the addition and subsequent formation of a vertex of a $K^0_S$ with the $D^*$
and muon, resulting in $N_{D_{s1}}$ candidates.
To find the branching ratio, the following formula is used:
\begin{eqnarray}
{f(\bar{b} \rightarrow B^0_s) \cdot 
Br({B}^0_{s} \rightarrow
{D}_{s1}^- \mu^+ \nu_{\mu} {X}) \cdot} \nonumber \hspace{1.0in}\ \\ 
\cdot {Br(D_{s1}^- \rightarrow D^{*-} K^0_S) = 
Br(\bar{b} \rightarrow D^{*-} \mu^+ \nu_{\mu} X) \cdot
\frac{N_{D_{s1}}}{N_{D^{*}\mu}} \cdot} \nonumber \\
{\frac{\epsilon(\bar{b} \rightarrow D^* \mu)}
{\epsilon(B^0_s \rightarrow D_{s1} \mu \rightarrow D^* \mu)}
\cdot 
\frac{1}{\epsilon_{K^0_S}} \thinspace \thinspace.} \hspace{1.3in}
\label{mastereq}
\end{eqnarray}
The input
$f(\bar{b} \rightarrow B^0_s)$~\cite{PDG}
is the
fraction of decays where a $b$ quark will hadronize to a
$B^0_s$ hadron. 
$\epsilon_{K^0_S}$ is the efficiency in
the signal decay channel to reconstruct and make a vertex with a $K^0_S$ 
to form
a $D_{s1}(2536)$, given that a $D^*$ and a muon have already been
reconstructed. 
Later we will identify the ratio of efficiencies as
$R^{{\mathrm{gen}}}_{D^*} =
 \epsilon(B^0_s \rightarrow D_{s1} \mu \rightarrow D^* \mu)/
 \epsilon(\bar{b} \rightarrow D^* \mu)$.

The D0\ detector~\cite{detect} and
following analysis~\cite{jason_thesis} are described in 
more detail elsewhere.
The main elements relevant to this analysis are the silicon
microstrip tracker (SMT),  
central fiber tracker (CFT), and muon detector systems.

This measurement uses a large data sample, 
corresponding to approximately 1.3~fb$^{-1}$ of integrated luminosity
collected by the D0 detector between April 2002 and March 2006.
Events were reconstructed using the standard D0 software 
suite.
To avoid
lifetime biases compared to the MC simulation, the small fraction of events
were removed that entered the sample only via triggers that included
requirements on impact parameters of tracks.


To evaluate signal mass resolution and  efficiencies, 
Monte Carlo (MC) simulated samples were generated for signal 
and background. 
The standard D0 simulation and event reconstruction chain was used.
Events were generated with the
{\sc pythia} generator~\cite{pythia} and decay chains of heavy
hadrons were simulated with the
{\sc evtgen} decay package~\cite{evtgen}.
The detector response was modeled by
{\sc geant}~\cite{geant}.
Two background MC samples were also generated: 
a $c\bar{c}$ sample, and
an inclusive 
$b$-quark sample containing all $b$ hadron species with forced
semileptonic decays to a muon.
In both cases, all events containing both a 
$D^{*}$ and a muon were retained.

\begin{figure}[htb!]\begin{center}
\centerline{\includegraphics[scale=0.32]{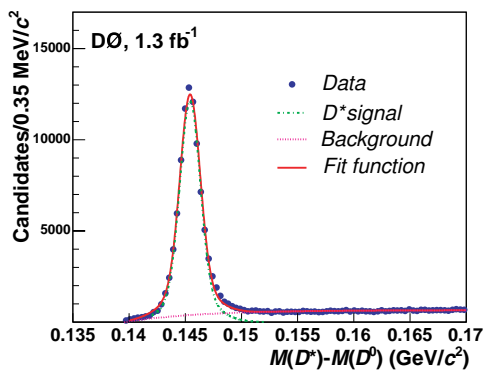}}
\caption{The mass difference $M(D^*) - M(D^0)$
  for events with $1.8 < M(D^0) < 1.95$ GeV/$c$$^2$ and
  an associated muon.
The number $N_{D^* \mu}$
was defined as the number of 
signal events in the mass difference range of 0.142--0.149 GeV/$c^2$. 
}
\label{fig:dst}
\end{center}
\end{figure}

$B$ mesons were first 
selected using their semileptonic decays, 
$B \rightarrow D^{*-} \mu^+ X$.
At this point in the selection, the $D^* + \mu$ sample 
is dominated by $B^0_d \rightarrow D^{*-} \mu^+ \nu_{\mu} X$ decays.
%
%
For this
analysis, muons were required to have hits in more than one 
muon layer, 
to have an associated track in 
the central tracking system,
and to have transverse momentum $p_T^{\mu} > 2$ GeV/$c$, 
pseudorapidity $|\eta^\mu|<2$, and total
momentum $p^{\mu} > 3$ GeV/$c$. 
%
 Two oppositely charged tracks with 
$p_T > 0.7$ GeV/$c$ and $|\eta|<2$ 
were required 
to form a common \adzero~vertex 
which were then combined with a muon 
candidate to form a common decay point following 
the procedure described in Ref.~\cite{dt}. 
For each \adzero$\mu^+$~candidate, 
an additional soft pion was searched for 
with charge opposite to the charge of the muon and $p_T > 0.18$ GeV/$c$. 
The ${K}^{-}$ and $\pi^+$ from the decay of the ${D}^{0}$
were both required to have more than five CFT hits.
To reduce the contribution from prompt $c\bar{c}$ production, 
a requirement was made on the transverse decay length, $L_{xy}$,
significance of the $D^* \mu$ vertex of 
$L_{xy}/\sigma(L_{xy}) > 1$.  
After these cuts, the total number of \dst 
candidates in the mass difference, $M(D^{*})-M(D^{0})$, 
peak of Fig.~\ref{fig:dst} 
is $N_{D^* \mu} = 87506 \pm 496$~(stat).

\ds $\thinspace$ candidates were formed by combining a $D^*$ candidate
with a $K_S^0$.  $D^*$ candidates were first selected 
by requiring the mass difference
$M( D^{*} ) - M( {D}^{0} )$ to be 
in the range  0.142--0.149 GeV/$c^2$.
The two tracks  from the 
decay of the $K_S^0$ were required to have opposite charge and to
have more than five hits in the CFT detector.
The $p_T$ of the $K_S^0$ was
required to be greater than 1~GeV/$c$ to reduce the contribution of
background $K^0_S$ mesons from fragmentation. A vertex was then formed
using the reconstructed $K^0_S$ and the $D^*$ candidate of the event.
The decay length of the $K^0_S$ was required to be greater than 0.5~cm.
To compute the $D_{s1}^{\pm}(2536)$ invariant mass, 
a mass constraint was applied using
the known $D^{*\pm}$  mass~\cite{PDG} instead 
of the measured invariant mass of the $K \pi \pi$ system.
Finally, the invariant mass of the
reconstructed $D_{s1}^{\pm}(2536)$ and muon was required to be less than the 
mass
of the $B^0_s$ meson~\cite{PDG}.

\begin{figure}[t]\begin{center}
\centerline{\includegraphics[scale=0.32]{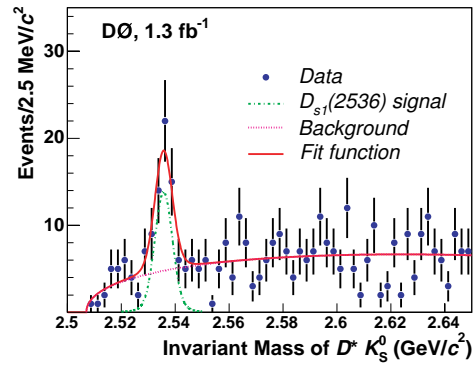}}
\caption{Invariant mass of $D^{*} K^0_S$ with an associated muon.  Shown is the result of the
  fit of the $D^{*} K^0_S$ mass with 
the function described in the text.}
\label{fig:mdstst_fin}
\end{center}
\end{figure}

The signal model employed for the fit 
to the $D^* K^0_S$ invariant mass spectrum was a relativistic Breit-Wigner 
convoluted with a Gaussian function, with
the resonance width fixed to the value $1.03 \pm 0.05 \thinspace \mathrm{(stat)} \pm 0.12 \thinspace \mathrm{(syst)} \thinspace \mathrm{MeV}/c^2$ measured
by the BaBar Collaboration~\cite{babarconf} 
and a Gaussian width determined to be $2.8$~MeV/$c^2$ from MC 
simulation of the signal.  The MC width value was scaled up by 
a factor of $1.10 \pm 0.10$ to account for differences between data and MC resolution estimates.
The unbinned likelihood fit used an 
exponential function plus a first-order polynomial 
to model the background with a threshold
cutoff of $M(D^*) + M(K^0_S)$.
The fit, shown in Fig.~\ref{fig:mdstst_fin}, 
gives a central value for the mass peak of 
$2535.7 \pm 0.7 \thinspace$(stat)~MeV/$c^{2}$, 
a yield of $N_{D_{s1}} = 45.9 \pm 9.1 \thinspace {\mathrm{(stat)}}$ events,
and a significance of $6.1 \sigma$ for the background to fluctuate up 
to or above the observed number of signal events. 

The efficiencies used in Eq.~\ref{mastereq} 
are estimated using the MC simulation,
after implementing suitable correction factors to ensure proper modeling
of the underlying  $b$-hadron $p_T$ spectrum, as well as trigger effects.
An event-by-event weight,
applied as a function of the generated $p_T$ of the
$B_s$, was determined by comparing the generated $p_T(B)$ in MC with the
$p_T$ distribution of fully reconstructed
$B^+ \rightarrow J/\psi K^+$ candidates in data collected primarily with
a dimuon trigger~\cite{Ralf}.  
Most events for this analysis were recorded using single muon
triggers, and an additional weight was applied as a function of
$p_T(\mu)$ to further improve the simulation of trigger effects.
Reweighted MC events were used in the determination of efficiencies 
described below, and indicated uncertainties are due to MC statistics.

Using the MC sample of inclusive $\bar{b} \rightarrow D^* \mu X$ events,
specific major decay modes were identified. Efficiencies for
each of these decay modes to pass the $D^*  \mu$ selection,
including the efficiency to reconstruct the soft pion from the 
$D^*$, were then determined.
The predicted fraction $F_i$ of each channel contributing to
the $D^*  \mu$ sample before further cuts was found following
a procedure similar to that given in Ref.~\cite{ratio_bdmixing}. 
The efficiency $\epsilon_i$ for each channel 
was found and a weighted sum was calculated, 
giving an estimated total efficiency for reconstruction of
$\epsilon(\bar{b} \rightarrow D^* \mu) = (5.88 \pm 0.80)\%$,
where the uncertainty is dominated by the MC statistics used
to find $\epsilon_i$, and uncertainties on external 
inputs~\cite{PDG} used to estimate $F_i$.
Applying the same cuts for reconstructing the $D^*  \mu$ for
the signal channel, the efficiency
$\epsilon(B^0_s \rightarrow D_{s1} \mu \rightarrow D^* \mu)
= (3.20 \pm 0.02)\%$,
results in a ratio
of efficiencies of  $R^{{\mathrm{gen}}}_{D^*} = 0.547 \pm 0.075$.

The signal MC sample was used to determine 
the efficiency to reconstruct
$D_{s1}^{-}(2536)  \rightarrow D^{*-} K^0_S$ 
given a reconstructed $D^* \mu$ as a starting point.
This efficiency is hence effectively that of reconstructing
a $K^0_S \rightarrow \pi^+ \pi^-$ and forming a vertex with the 
$D^* \mu$, and includes the branching ratio
$Br(K^0_S \rightarrow \pi^+ \pi^-)$~\cite{PDG} for
ease of use in calculating the branching ratio product.
The reconstruction efficiency 
was found to be 
$\epsilon_{K^0_S} = (10.3 \pm 0.4$)\% where the  uncertainty is due to
MC statistics.




The process $c\bar{c} \rightarrow D^{*-} \mu^+ \nu_{\mu} X$ 
can contribute to $N_{D^{*} \mu}$ since a $D^*$ meson can come from the hadronization of the 
$\bar{c}$ quark, 
and the muon can come from the semileptonic decay of the hadron
containing the $c$ quark.  
To determine the number of events in our signal
reconstructed from a prompt $D^{*}$, 
a comparison was made of the decay length significance distribution observed
in the data with the same distribution
predicted by MC for $b \rightarrow D^* \mu X$ and
any excess at shorter significances was interpreted as $c\bar{c}$ contribution.
For the decay length significance cut used in the analysis, 
$L_{xy}/{\sigma(L_{xy})} > 1$, 
the fraction of $N_{D^* \mu}$ from $c\bar{c}$ 
production was estimated to be $(3.9 \pm 2.5)\%$.
A check using a prompt $c\bar{c}$ MC sample results in a 
consistent estimate.
The value of $N_{D^* \mu}$ was corrected downward accordingly.  

The contribution from
$c\bar{c}$ production to $N_{D_{s1}}$ where one charm quark 
hadronizes directly to a $D_{s1}(2536)$ and the other decays directly 
to a muon was estimated to be negligible using  
relative production ratios and 
spin-counting arguments~\cite{charm_CDF_Peskin}.

Systematic uncertainties for the branching ratio product are 
summarized in Table~\ref{tabsyst} and discussed below.
The uncertainty in the normalizing branching ratio~\cite{PDG} 
$Br(\bar{b} \rightarrow D^* \mu X)$
was taken as a systematic uncertainty. 
For determining $N_{D^* \mu}$, the signal and background model
parameters
were varied
in a correlated fashion and a systematic uncertainty was assigned. 
The estimated $c\bar{c}$ production contribution 
was varied by the
indicated uncertainty. In the determination of $N_{D_{s1}}$, the functional 
forms of the signal and background models
were varied in a number of ways
to determine the sensitivity of the candidate yield. 
In addition, the scaling of the widths
was varied by $\pm 10\%$ 
to check the sensitivity to uncertainty
in mass resolution.

 By comparing the $p_T(\mu)$ distribution for the signal
using the default ISGW2 decay model~\cite{ISGW} to the HQET semileptonic
decay model~\cite{evtgen},
a weighting factor was found and applied to the fully simulated signal MC
events,
and the efficiency
determined again. The difference observed was assigned
as a contribution to the systematic uncertainty of 
$\epsilon_{K^0_S}$ and $R^{{\mathrm{gen}}}_{D^*}$. 

When estimating  $\epsilon_{K^0_S}$, the uncertainty due to
modeling of the $b$ hadron $p_T$ spectrum was derived
by using an alternate weighting technique. 
The cuts on the $p_T$ and decay length of the $K_{S}^{0}$ were varied 
and a systematic uncertainty on the efficiency due to this source was also 
assigned.
Discrepancies in track reconstruction efficiencies between data and MC 
in low-$p_T$ tracks were accounted for by assigning a systematic 
uncertainty to each of the pion tracks in the 
$K_{S}^{0}$ reconstruction~\cite{andrzej, narrow}.

The uncertainty in $R^{{\mathrm{gen}}}_{D^*}$ is
due to a combination of MC statistics and uncertainties
in PDG branching ratio values and production fractions, 
$f(\bar{b} \rightarrow b 
\thinspace {\mathrm{hadron}})$. The uncorrelated
systematic uncertainty is given in Table~\ref{tabsyst}.


The estimated systematic uncertainties were added in quadrature to obtain a total estimated systematic uncertainty on the branching ratio product of 16.8\%.
The branching ratio product was determined  to be:
\begin{eqnarray}
  f(\bar{b} \rightarrow B^0_s) \cdot
 Br({B}^0_{s} \rightarrow
 {D}_{s1}^- \mu^+ \nu_{\mu} {X}) 
 \cdot Br(D_{s1}^- \rightarrow D^{*-} K^0_S)=  \nonumber\\
= [2.66 \pm 0.52
\thinspace {\mathrm{(stat)}}
\pm 0.45 \thinspace {\mathrm{(syst)}}] \times 10^{-4}.  \nonumber
\end{eqnarray}

\begin{table}[htb]\begin{center}
\caption{\label{tabsyst}
Estimated systematic uncertainties.}
\begin{tabular}{lc}
\hline
\hline
Source & Systematic uncertainty \\
\hline
$Br(\bar{b} \rightarrow D^* \mu X)$  & 6.9\% \\
$N_{D^* \mu}$ & 2.9\% \\
$N_{D_{s1}}$ & 5.5\% \\
$\epsilon_{K^0_S}$ & 11.0\%\\
$R^{{\mathrm{gen}}}_{D^*}$ & 8.6\% \\ 
\hline
Total & 16.8\% \\
\hline
\hline
\end{tabular}
\end{center}
\end{table}

 

To assess the systematic uncertainty on the mass measurement, 
the same variations of the $D_{s1}(2536)$ mass signal model, as
well as background functional form, were applied  as described above.  
The mass values used for
the mass constraints on the decay products were varied within their PDG
uncertainties and were also set to the D0 central fit values. 
Ensemble tests indicated that the statistical error is correct.  
From the observed
variations, a total
systematic mass uncertainty of $0.5$~MeV/$c^2$ was taken, for a mass
measurement of: 
$$ m(D_{s1}) = 2535.7 \pm 0.6 \thinspace {\mathrm{(stat)}} 
                       \pm 0.5 \thinspace {\mathrm{(syst) \thinspace
		       MeV}}/c^2.
		       \nonumber
$$
This measured mass value 
is in good 
agreement with 
the PDG average value
of $2535.34 \pm 0.31$~MeV/$c^2$~\cite{PDG}.


To allow comparison of this measurement to 
theoretical predictions, the semileptonic branching ratio
alone
as shown in Table~\ref{tab_comp}
is extracted by taking the hadronization fraction
into $B^0_s$ as
$f(\bar{b} \rightarrow B^0_s) = 0.103 \pm 0.014$~\cite{PDG} and also
assuming that $Br(D_{s1}(2536) \rightarrow 
D^* K^0_S) = 0.25$~\cite{evtgen}. 
This is the first experimental measurement of this
 semileptonic branching ratio and is compared to a number of theoretical 
predictions~\cite{pred1,pred2,pred3} of the 
exclusive rate in Table~\ref{tab_comp}.
The systematic uncertainty on this quantity is as described earlier, 
and the error labeled ``(prod. frac.)" is due to the
current uncertainty on 
$f(\bar{b} \rightarrow B^0_s)$. 
The first two theoretical predictions include relativistic and $1/m_Q$
corrections, while the third does not. The result is found to be 
consistent within uncertainties with the first two theoretical 
predictions, and demonstrates the need for such corrections. 

\begin{table}[htb]
\caption{\label{tab_comp} Experimental measurement
compared with various theoretical predictions.}
\begin{tabular}{lc}
\hline
\hline
Source & $Br(B^0_s \rightarrow D_{s1}^-(2536) \mu^+ \nu_{\mu} X)$ \\
\hline
  This result  & 
$[1.03 \pm 0.20 \thinspace {\mathrm{(stat)}}\pm 0.17 \thinspace {\mathrm{(syst)}}$\\ 
       & 
      $ \pm \thinspace 0.14 
      \thinspace {\mathrm{(prod. \thinspace frac.)}}]$\% \\ 
\hline
Theoretical Predictions   & $Br(B^0_s \rightarrow D_{s1}^-(2536) \mu^+ 
\nu_{\mu})$ \\
\hline
ISGW2~\cite{pred1} & $(0.53 \pm 0.27)$\% \\
Relativistic Quark Model \& \\ 
\thinspace \thinspace \thinspace $1/m_Q$ corrections~\cite{pred2}   & $(1.06 \pm 0.16)$\% \\
Non-rel. HQET and ISGW~\cite{pred3} & 0.195\% \\
\hline
\hline
\end{tabular}
\end{table}

In summary, using 1.3 fb$^{-1}$ of integrated luminosity collected with the
D0 detector, 
a first measurement of the semileptonic $B^0_s$ decay into the narrow 
$D_{s1}^{\pm}(2536)$ state has been made and compared with theory.  
In addition, the mass of the $D_{s1}^{\pm}(2536)$ was measured 
and found to be in good agreement with the PDG value.

%
We thank the staffs at Fermilab and collaborating institutions, 
and acknowledge support from the 
DOE and NSF (USA);
CEA and CNRS/IN2P3 (France);
FASI, Rosatom and RFBR (Russia);
CAPES, CNPq, FAPERJ, FAPESP and FUNDUNESP (Brazil);
DAE and DST (India);
Colciencias (Colombia);
CONACyT (Mexico);
KRF and KOSEF (Korea);
CONICET and UBACyT (Argentina);
FOM (The Netherlands);
Science and Technology Facilities Council (United Kingdom);
MSMT and GACR (Czech Republic);
CRC Program, CFI, NSERC and WestGrid Project (Canada);
BMBF and DFG (Germany);
SFI (Ireland);
The Swedish Research Council (Sweden);
CAS and CNSF (China);
Alexander von Humboldt Foundation;
and the Marie Curie Program.
%

\end{document}